\documentstyle[prl,aps,twocolumn,floats]{revtex}

\newcommand\journal[5]{#1, #2 {\bf #3} (#4) #5}

\newcommand\NPB{Nucl. Phys.}
\newcommand\PR{Phys. Rev.}

\newcommand{\tr}[1]{\:{\rm Tr}\,#1}
\def\e{{\, e}\,}

\newcommand{\vgr}{{\vec \nabla}}
\newcommand{\vx}{{\vec x}}
\def\be{\begin{equation}}
\def\ee{\end{equation}}
\def\bea{\begin{eqnarray}}
\def\eea{\end{eqnarray}}
\def\nn{\nonumber}

\def\v{\varphi}
\def\s{\sigma}
\def\vcl{\varphi_{\rm cl}}
\newcommand{\no}[1]{:\!#1\!:}
\def\la{\left\langle}
\def\ra{\right\rangle}
\def\d{\partial}
\def\se{S_{\rm eff}}

\begin{document}
\twocolumn[\hsize\textwidth\columnwidth\hsize\csname
@twocolumnfalse\endcsname
{\hspace{150mm}ITEP-TH-48/98}
\title{{\Large Universality in Effective Strings}}
\author{\bf S. Jaimungal,
G. W. Semenoff and
K. Zarembo}
\address{{\it Department of Physics and Astronomy,
University of British Columbia \\
6224 Agricultural Road,
Vancouver, British Columbia V6T 1Z1}}
\date{\today}
\maketitle
\begin{abstract}
 We
demonstrate that, due to the finite thickness of domain walls,
and the consequent ambiguity in defining their locations, 
the effective string description obtained by integrating out bulk
degrees of freedom contains ambiguities in the coefficients of the
various geometric terms. The only term with unambiguous coefficient is
the zeroth order Nambu-Goto term. We argue that fermionic ghost fields which implement gauge-fixing act to balance
these ambiguities. The renormalized string tension, obtained after
integrating out both bulk and world-sheet degrees of freedom, can be
defined in a scheme independent manner; and we compute the explicit
finite expressions, to one-loop, for the case of compact quantum
electrodynamics and $\v^4$ theory.
\end{abstract}
~]

A long standing problem in the physics of interfaces in three dimensional
systems is to describe  the interface dynamics as a theory of fluctuating
surfaces analogous to an effective Euclidean string theory
\cite{jas84,zia85,Mun98}. The interface surface can be 
interpreted as the world-sheet of an effective string in
three-dimensions, while the two phases on either side of the interface
represent the vacuum expectation values of a fundamental field. For example in the Ising model, the field is the spin operator, and the
interface is the set of links about which the spin changes sign. An
effective string action of this surface has a typical geometric
expansion which begins with the Nambu-Goto term (the area), next the
extrinsic curvature and then higher order curvature corrections. It is
usually argued that the higher order corrections are irrelevant as far
as the low energy dynamics is concerned, and perturbation theory
consisting in keeping only the area term and extrinsic curvature is
valid. However, we will demonstrate that such an ansatz is in fact
ambiguous from the outset.  In particular, we will demonstrate that
coefficients of the higher order curvatures are completely arbitrary,
and depend upon the precise prescription implemented in defining the
position of the interface. Such an ambiguity is a direct consequence
of the finite thickness of the interface region. In spite of this
ambiguity an effective string description is possible if one includes
in the action fermionic degrees of freedom which reinstate the scheme
independent nature of the fundamental action.

In this work we concentrate on interfaces which occur in
three-dimensional compact quantum electrodynamics (QED) \cite{pol77}. 
The analysis can be easily applied to any other field
theory in three-dimensions which has a soliton solution to the
classical equations of motion. In compact QED, monopole instantons cause 
the electric fields between two charged particles to
form a flux tube, the potential between electric charges
grows linearly with the distance between them, and 
the charges are confined \cite{pol77}. This picture
of confinement is, however, a purely classical one. In the full
quantum theory the string of electric flux, along with the magnetic
fields in the bulk, are not rigid but rather fluctuate.

Compact QED can be regarded as the low-energy effective theory for the
Georgi-Glashow model where the $SU(2)$ symmetry is spontaneously
broken to $U(1)$.  The monopoles appearing in the broken gauge theory
are the classical solutions of the original Higgs model which have
finite Euclidean action \cite{tp74}. The collection of monopoles
behave as a gas of charged particles interacting through a Coulomb
force. Since the charges are of order $\sqrt{4\pi}/g$, where $g$ is
the gauge coupling, the monopole configurations can be treated using a semi-classical approximation in the limit of weak gauge
coupling. The system then reduces to the classical thermodynamics of a
Coulomb gas and the partition function is
a Sine-Gordon (SG) theory
\cite{pol77}:
\bea
Z &=& \int[d\v]\,\exp\left\{-\frac{g^2}{32\pi^2}\,
\int d^3x\,\left[ (\d\v)^2 \right.\right. \nn\\
&&\left.\left.\hspace{40mm} -2m^2\no{\cos\v}\right]\right\}
\label{defz}
\eea
The monopole density is the
coefficient of the cosine interaction, $\zeta=g^2m^2/32\pi^2$.  The
Grand canonical partition function for the monopoles is recovered 
by expanding in powers of $\zeta$ and
performing the functional integral over $\v$.  In the presence of the
monopole solution the photon becomes massive and Wilson loop
correlators obey an area law \cite{pol77}. The relevant order
parameter in the original gauge theory (compact QED) is the vacuum
expectation value of the Wilson loop. There is a natural mapping
between this correlator and the following correlation function in the
SG model \cite{pol77,pol97}:
\bea\label{wls}
W(C) &\equiv&
\la\exp\left(\frac{i}{2}\oint_C dA \right)
\ra_{\rm QED}\nn\\
&=& \la \exp\left(\frac{g^2}{16\pi}
\int_S \star {\rm d} \v \right)\ra_{\rm SG} \label{WilsonLoop}
\eea
Here $S$ is an arbitrary surface bounded by the contour $C$.  
The result can be shown to be independent of the choice of
this surface.  We are interested in the behaviour of 
(\ref{wls}) for large loops.  The contour $C$ will
be assumed to lie at infinity in the $x_3=0$ plane. In this case, it
is possible to reformulate the problem. The operator on the right hand
side of (\ref{wls}) introduces a source for the SG field or,
equivalently, one may assume that $\v$ experiences a jump of magnitude
$2\pi$ across the surface $S$. Since the potential is periodic in
$\v$, this jump can be eliminated by shifting $\v$ on one side of the
surface by $2\pi$. This shift renders the field continuous, however,
it changes the boundary conditions as
$x_3\rightarrow+\infty$. Consequently, performing such a shift on $\v$
reduces (\ref{WilsonLoop}) to an evaluation of the path integral
\ref{defz} with the boundary conditions $\v\rightarrow 0$ as
$x_3\rightarrow -\infty$ and $\v\rightarrow 2\pi$ as $x_3\rightarrow
+\infty$. These boundary conditions are precisely what is required for
field configurations in the presence of a domain wall. The domain wall
in this case describes a world sheet of a string of electric flux
created by charges at infinity.

The first step in obtaining an effective string action is to obtain
the classical solution satisfying the appropriate boundary
conditions. The solution is the SG soliton:
\be\label{sol}
\vcl(x)= 4 \arctan\,\e^{m x_3},
\ee
which corresponds to a domain wall in the $x_3=0$ plane. The
position of this domain wall is in fact ambiguous. Conventionally, its
position is given by the surface on which $\v=\pi$
\cite{pol97,sky62}. However, other definitions are also possible, for
example, the surface on which the energy density is maximal
\cite{GS75,GJST75}. These definitions, although agreeing at the
classical level, do not agree once quantum corrections are
included. Nevertheless, they will yield the same result within an
order of $m^{-1}$. This uncertainty is due to the finite thickness of
the domain wall and we will argue that this ambiguity translates into
the non-universality of the world-sheet action.

Upon inserting (\ref{sol}) as a classical background field in 
the functional integral over
$\v$ in (\ref{defz}), the collective coordinate method can be used
to separate the integral over fluctuations 
into an integral over the domain wall position,
which we describe by a height function $f(x_1,x_2)$, and over the
field fluctuations in the bulk:
\be\label{dw}
Z = \int [d\v][df] \Delta_{\rm FP}[\v]
\delta\left[K[f,\phi]\right] \e^{-S_{\rm SG}[\v]}
\ee
where $K[f,\phi]=\int dx_3 K(\vx,x_3- f)\v(x) - \pi$ and
$\Delta_{\rm FP}[\v]=\delta K[f,\phi]/\delta f$ 
is the Faddeev-Popov determinant, $\vx$
is $(x_1,x_2)$ and $S_{\rm SG}$ is
the same as in (\ref{defz}). A particular definition of the domain wall
position corresponds to choosing a kernel $K$. 
There is no unique choice of this function.  The
definition of \cite{pol97,sky62} corresponds to $K=\delta(x_3-f)$,
while the standard collective coordinate method\cite{GS75,GJST75}, in which the
fluctuations of the domain wall are associated with quasi-zero modes
in the background of the classical solution (\ref{sol}), corresponds to
$K=\vcl'(x_3-f)$. 

Integrating over $\v$ in (\ref{dw}) yields an effective action for the coordinate $f(\vx)$
of the domain wall. In the semi-classical approximation $\v$ is
replaced by its classical solution. The effective action is then given
by,
\be
S_{\rm eff.} = \sigma_0 \int \left( 1 + \frac{1}{2} (\vgr f)^2
\right) + {\cal O}\left ( (\vgr f)^2 \right)
\ee
where the string tension is determined by the mass of the SG soliton
\cite{pol77,sny83}:
$$
\s_0
= \frac{g^2}{32\pi^2}\int \left( (\d_3\vcl)^2 + 2 m^2 cos(\vcl)
\right) dx_3 = \frac{g^2m}{2\pi^2} $$ In $S_{\rm eff.}$ all higher
order corrections, coming from tree level diagrams, were
ignored. However, if terms which would contribute to higher curvature
corrections are ignored, it is possible to re-sum an infinite subset
of the terms which were previously ignored.  This can be achieved by
considering a domain wall solution which is curved, unfortunately,
such configurations are not solutions of the SG equations of
motion. They can, on the other-hand, be considered as a constraint
solution \cite{aff81} -- the solution of the equations of motion with
a source term proportional to the argument of the delta function in
(\ref{dw}). This can be included in the action by introducing a
Lagrange multiplier field.  For slowly varying $f$ this equation can
be solved by perturbation theory in the derivatives of $f$.  If only
the first derivatives of $f$ are taken into account (i.e.  ignoring
higher curvature corrections), the solution can be constructed by the
following simple arguments \cite{zia85}. When one neglects higher
derivatives this implies that $f$ is a linear function of its
arguments, which corresponds to the plane domain wall rotated through
an angle $\theta$ with $\tan\theta=\sqrt{G}$ where $G\equiv 1+(\vgr
f)^2$. The classical solution in this case is obtained from
(\ref{sol}) by rotation: $\v(x)=\vcl((x_3-f)/\sqrt{G})$. This solution
is exact in the approximation of constant $\vgr f$, since no source
term is required to produce it. Consequently, the Lagrange
multiplier appears only in the next order of the derivative expansion
and is proportional to $\vgr^2 f$. The effective action with the above
solution is equal to the area of the domain wall. The rotated wall
 area element contains the factor $1/\cos\theta=\sqrt{G}$, and the
re-summed effective action is therefore,
\be\label{ng}
\se=\s_0\int d^2\vx\,\sqrt{G\left(\vx\right)}.
\ee
Since $G$ is the determinant of the induced metric on the
string world sheet we have obtained the Nambu-Goto action.  

The next term in the derivative expansion will be of order of the
Lagrange multiplier squared, that is $(\vgr^2 f)^2$, and will depend
on the choice of the constraint. This arbitrariness leads to an
ambiguity in the coefficient of this term, which in the covariant
description corresponds to the extrinsic curvature squared. It appears
that the area term (\ref{ng}) is the only universal part of the
effective action.  This is not surprising, since the parameter of the
derivative expansion is $k/m$, where $k$ is a momentum of the
excitation on the string world sheet. Higher derivative terms become
important when $k\sim m$, i.e., when the wave-length of the excitation
is of the same order of the thickness of the domain wall. Such
excitations are of course indistinguishable from the fluctuations of
the SG field in the bulk. As such, including such fluctuations in the
effective string action leads to ambiguities. Of course, the
computation of any physical quantity must be invariant under any
choice of the constraint. It is the Faddeev-Popov determinant which
cancels the ambiguities arising in the bosonic sector of the theory -
the full effective action contains the fermionic ghosts coming from
the determinant.

The universality of the Nambu-Goto term stems from the rotational
invariance of the original model. This holds even when quantum
corrections are included as long as the constraint respects rotational
symmetry. Constraints which do not respect this symmetry would produce
actions that are not reparametrization invariant.  We have explicitly
checked that when the kernel in (\ref{dw}) is chosen to be
$K(\vx,x_3-f)=\vcl'(x_3-f)$, which is not rotationally invariant, the
one-loop correction to the constant and to the $(\vgr f)^2$ terms in
(\ref{ng}) disagree. 

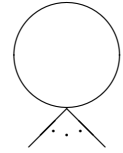
\begin{figure}
\setlength{\unitlength}{0.1in}
\begin{center}
\vskip 5mm
\begin{picture}(3,5)(0,8.2)
\put(2,12){\circle{10}}
\put(2,9.2){\line(1,-1){2}}
\put(2,9.2){\line(-1,-1){2}}
\put(1.3,8){\circle*{.1}}
\put(2,7.8){\circle*{.1}}
\put(2.7,8){\circle*{.1}}
\end{picture}
\vskip 5mm
\caption{Tadpole diagrams to be subtracted in the normal ordering 
prescription.\label{bubble}}
\end{center}
\vskip-5mm
\end{figure}

Let us now consider the quantum corrections.  The string
tension gains quantum corrections from fluctuations of the domain wall
and of the field $\v$ in the bulk. To study how these two 
corrections are correlated we calculate the one-loop corrections to
the string tension. The parameter of the loop expansion, $m/g^2$, is
small since $m^2$ is proportional to the monopole density.  These
corrections will be computed in the background field method starting
from the classical solution (\ref{sol}).  The one-loop corrections in 3D
SG theory are linearly divergent, which leads to ambiguities in the
definition of dimensional quantities like the string tension. However,
the monopole gas representation implies the particular UV
regularization based on the normal ordering prescription. To implement
this prescription in the background field method, it is instructive to
first study the one-loop corrections to the general classical
solution. Expanding the quantum field as $\v=\vcl+\eta$ and
integrating out $\eta$, we obtain: $$ S^{\rm
bare}_1=\frac{1}{2}\,\tr\ln\left(\frac{-\d^2+m^2\cos\vcl}{-\d^2+m^2}
\right).
$$ The expansion of $S_1$ in the powers of $\vcl$ yields the usual
Feynman diagrams while the normal ordering prescription consists in
throwing out bubble diagrams (Fig. \ref{bubble}). Thus, to one-loop,
the normal ordering is implemented by adding the following
counter-term to the effective action:
\bea
S_1 &=& \frac{1}{2}\tr\ln\left(\frac{-\d^2+m^2\cos\vcl}{-\d^2+m^2}\right) \nn\\
&-&\frac{m^2}{2}\int d^3x(\cos\vcl-1)
\int\frac{d^3p}{(2\pi)^3} \frac{1}{p^2+m^2} \label{oneloop}
\eea
This expression is free of all UV divergences. Notice that since the
expansion of the renormalized effective action $S_1$ begins with
$\vcl^4$ terms, the mass of the photon does not acquire quantum
corrections at one loop.

The spectrum of the operator describing the excitations of the soliton
are well known. It consists of plane waves in the $x_1$ and $x_2$
directions; while the $x_3$ spectrum is gapped possessing a zero mode
and then a set of continuum states corresponding to the scattering of
plane waves off the potential given by the solitonic solution.
Inserting the analytic form of the spectrum into
(\ref{oneloop}) we find that the one-loop correction to the string
tension is,
\bea \label{dsigmasg}
\Delta \s_1
&=& \frac 1 2 \int \frac{d^3p}{(2\pi)^3} \ln ( p^2 + m^2) 
\nu(p_3)\nn\\
&+& \frac 1 2 \int  \frac{d^2 p}{(2\pi)^2} \ln (\vec{p}^{\,2} )
+ 2m \int \frac{d^3 p}{(2\pi)^3} \frac 1 {p^2+m^2} \nonumber\\
&=& - \frac {m^2}{4\pi}.
\eea
Here $\nu(p_3)$ is the difference of the density of scattering states
in the presence and absence of the kink, and can be obtained using
the exact eigenmodes of the linearized SG equation \cite{GJ75}:
$$\nu(p_3)=\int dx_3 \left(
\psi_{p_3}^*(x_3)\psi_{p_3}(x_3) - 1 \right)= -\;\frac
{2m}{p_3^2+m^2}$$ The first term in (\ref{dsigmasg})
corresponds to the trace over continuum states; the second
term corresponds to the quasi-zero mode contribution; and the last
term is the counter-term prescribed by normal ordering. The sum of
these three terms is UV finite, although each term diverges
individually. This leaves room for some ambiguity in the final finite
result depending on what regularization scheme is
implemented. However, if the sum of the terms is placed under one
single integral, so that the function being integrated over has finite
UV behaviour, the answer must be regularization independent. The
difficulty in implementing such a scheme is that the quasi-zero mode
is a two dimensional integral, while all others are three-dimensional. It is possible to integrate out the $p_3$ component of
the three-dimensional integrals first, which then leaves a single two
dimensional integral to perform. This is the scheme that was
used in (\ref{dsigmasg}).

Thus far, we have shown that after including one-loop effects the
string tension in the SG model is,
$$
\sigma =  \frac{g^2m}{2\pi^2} \left( 1 - \frac {\pi}{2}~\frac {m}{g^2} \right).
$$
We performed analogous calculation for $\v^4$ theory with the potential
$\lambda(\v^2-m^2/\lambda)^2/2$, the result is,
$$
\sigma = \frac 4 3 \frac {m^3}{\lambda} \left ( 1 - \frac {9}{32\pi} 
(4- \ln 3 ) \frac \lambda m \right ).  $$ As mentioned earlier,
different regularization schemes will lead to different final
answers. The authors of \cite{KO89} performed similar calculations on
the $\v^4$ theory using the zeta-function regularization. That
computation, however, did not include the integral over the quasi-zero
branch of the theory, and therefore corresponds to obtaining the
one-loop correction to the effective string tension rather than the
one-loop renormalized string tension. Excluding that branch still
leads to finite results in their regularization scheme since it is
insensitive to power like divergences. Unfortunately, it yields an
answer incompatible with the ansatz of first integrating out $p_3$ and
then performing the finite integrals. This is to be expected, since
the quasi-zero branch should and must contribute to the renormalized
string tension.

In the preceding, all modes were including in computing
the determinant. However, it is possible to integrate out only the
scattering states, {\it i.e.} to omit the second term in
(\ref{dsigmasg}), and obtain an effective action for the quasi-zero
branch (as in \cite{KO89}).  This branch contains the modes
responsible for shifting the surface in the $x_3$ direction.  
Omitting the second term in \ref{dsigmasg} is equivalent to
keeping $f$ fixed and integrating out only bulk modes.  The one-loop
correction with only bulk modes included is badly UV divergent and is
not regularization independent. In the naive cut-off regularization
the effective string tension for the $f$ fields will be $\sim
\Lambda^2 \ln(\Lambda^2 /m^2)$ as can be easily checked from
(\ref{dsigmasg}). Of course, hard modes (with $k\gg m$) of the field
$f$ cancel this divergence, rendering the physical string tension
finite.  The implication is that the one-loop renormalization of the
string tension is known precisely, even though all the interaction
terms are not known.

To summarize we have argued that the accuracy of the macroscopic
description of the confining string in compact QED is limited by the
fact that the string is not infinitely thin. As a consequence the
higher-derivative terms in the world-sheet action are not
universal. Formally, it is possible to obtain an effective string
theory action by integrating out $\v$ in (\ref{dw}) exactly. However,
the world-sheet action will be scheme-dependent and must be
supplemented by the fermionic ghosts coming from the constraints. In
addition, the action contains rather peculiar divergences and the
finite physical quantities obtained from such an effective action
appear only after delicate cancellations between these divergences and
the contribution of hard modes of the string coordinates.

It is worth mentioning that the derivation of the effective domain
wall action by collective coordinate method in one dimension lower
would lead to essentially the same conclusions. In 2D theory, which
can be thought of as a high-temperature reduction of 3D one
\cite{az97}, domain walls correspond to soliton paths. Solitons in 2D
SG theory are known to be described by a local field theory, the
Thirring model \cite{Coleman} and soliton operators \cite{Mandelstam}
look very much like the dimensional reduction of the Wilson loops
(\ref{wls}) \cite{az97}. Where the only difference is a local factor
rendering the solitons fermionic. Fermion propogators have a well
defined sum-over-path representation where the action is the
supersymmetrized length of the world-line. Nevertheless, at weak SG
coupling, solitons cannot be described by the world-line theory,
since the four-fermion interaction in the Thirring model, which
corresponds to a contact interaction in the sum-over-path picture,
becomes infinitely strong.

We are grateful to K.~Selivanov and A.~Zhitnitsky for
discussions. This work is supported in part by NSERC of Canada, a NATO
Science Fellowship and, in part, by RFFI grant 98-01-00327.

\end{document}